\documentclass[11pt,letterpaper]{article}
\pdfoutput=1
\usepackage{jheppub}
\usepackage{bbm}
\usepackage{mathrsfs}
\usepackage{slashed}
\usepackage{caption}
\usepackage{epstopdf}
\usepackage[normalem]{ulem}
\usepackage[bottom]{footmisc}
\usepackage{subcaption}
\usepackage{bbold}
\usepackage{titlesec}
\usepackage{threeparttable}
\usepackage{booktabs}
\usepackage{changepage}
\usepackage[utf8]{inputenc}
\usepackage{dsfont}

\usepackage{grffile}
 
\usepackage{graphicx}  
\usepackage{dcolumn}   
\usepackage{bm}        
\usepackage{amssymb}   
\usepackage{setspace}
\usepackage{amsmath, amssymb, setspace}
\usepackage{array}
\usepackage{booktabs}
\usepackage{caption}
\usepackage{indentfirst}
\usepackage{float}
\usepackage{lmodern}
\usepackage{multirow}
\usepackage{soul}

\usepackage{braket}
\usepackage{comment}

\usepackage[draft]{pgf}

\usepackage{adjustbox} 
\newcommand{\dd}[2]{\frac{d #1}{d #2}}

\newlength{\myimageoversize}
\newsavebox{\myimage}

\usepackage[most]{tcolorbox}

\usepackage{empheq}

\tcbset{colback=white!10!white, colframe=black!50!black, 
        highlight math style= {enhanced, 
            colframe=black,colback=white!10!white,boxsep=0pt}
        }

\titleformat{\subsubsection}
 {\normalfont\fontsize{12}{17}\itshape}{\thesubsubsection}{1em}{}
 
\title{ \huge{Analytic Description of Primordial Black Hole Formation from Scalar Field Fragmentation}}

\author[a,b]{Eric Cotner,}
\author[a,c]{Alexander Kusenko,}
\author[c,d,e]{Misao Sasaki,}
\author[a]{Volodymyr Takhistov}

 \affiliation[a]{Department of Physics and Astronomy, University of California -- Los Angeles,\\
 Los Angeles, CA 90095-1547, USA}
 \affiliation[b]{Advanced Analytics Center of Excellence, American Tire Distributors, \\
Huntersville, NC 28078, USA}
 \affiliation[c]{Kavli Institute for the Physics and Mathematics of the Universe (WPI), UTIAS\\
The University of Tokyo, Kashiwa, Chiba 277-8583, Japan}
\affiliation[d]{Center for Gravitational Physics, Yukawa Institute for
Theoretical Physics,\\
Kyoto University, Kyoto 606-8502, Japan}
\affiliation[e]{Leung Center for Cosmology and Particle Astrophysics,
National Taiwan University,\\
Taipei 10617, Taiwan}

 \emailAdd{ecotner@ucla.edu}
 \emailAdd{kusenko@ucla.edu}
 \emailAdd{misao.sasaki@ipmu.jp}
 \emailAdd{vtakhist@physics.ucla.edu}
 
\abstract{
Primordial black hole (PBH) formation is a more generic phenomenon than was once thought.  The dynamics of a scalar field in inflationary universe can produce PBHs under mild assumptions regarding the scalar potential. In the early universe, light scalar fields develop large expectation values during inflation and subsequently relax to the minimum of the effective potential at a later time. During the relaxation process, an initially homogeneous scalar condensate can fragment into lumps via an instability similar to the gravitational (Jeans) instability, where the scalar self-interactions, rather than gravity, play the leading role.  The fragmentation of the scalar field into lumps (e.g. Q-balls or oscillons) creates matter composed of relatively few heavy ``particles", whose distribution is subject to significant fluctuations unconstrained by comic microwave background (CMB) observations and unrelated to the large-scale structure.  If this matter component comes to temporarily dominate the energy density before the scalar lumps decay, PBHs can be efficiently produced during the temporary matter-dominated era. We develop a general analytic framework for description of PBH formation in this class of models. We highlight the differences between the scalar fragmentation scenario and other  commonly considered PBH formation models. Given the existence of the Higgs field and the preponderance of scalar fields within supersymmetric and other models of new physics, PBHs constitute an appealing and plausible candidate for dark matter. 
 }

\begin{document}
\preprint{IPMU19-0063, YITP-19-31}
 \maketitle
\flushbottom

\section{Introduction}
	
Standard astrophysical black holes form as a result of stellar collapse at relatively recent cosmological times, after the onset of star formation. However, black holes could also form in the early universe. Such primordial black holes (PBHs) can account for all or part of the dark matter (DM)~(e.g.~\cite{Zeldovich:1967,Hawking:1971ei,Carr:1974nx,GarciaBellido:1996qt,Khlopov:2008qy,Frampton:2010sw,Kawasaki:2016pql,Cotner:2016cvr,Carr:2016drx,Inomata:2016rbd,Pi:2017gih,Inomata:2017okj,Garcia-Bellido:2017aan,Inoue:2017csr,Georg:2017mqk,Inomata:2017bwi,Kocsis:2017yty,Ando:2017veq,Sasaki:2018dmp,Carr:2018rid}). PBHs can be associated with a variety of astrophysical phenomena, including the recently discovered \cite{Abbott:2016blz,Abbott:2016nmj,Abbott:2017vtc} gravitational waves (e.g.~\cite{Nakamura:1997sm,Clesse:2015wea,Bird:2016dcv,Raidal:2017mfl,Eroshenko:2016hmn,Sasaki:2016jop,Clesse:2016ajp,Takhistov:2017bpt}), formation of supermassive black holes~\cite{Bean:2002kx,Kawasaki:2012kn,Clesse:2015wea} as well as  $r$-process nucleosynthesis~\cite{Fuller:2017uyd}, gamma-ray bursts and microquasars \cite{Takhistov:2017nmt} from compact star disruptions.  
	
In conventional PBH production mechanisms  (see e.g. Ref.~\cite{Carr:2016drx,Sasaki:2018dmp} for review) BHs form when density fluctuations are of $\mathcal{O}(1)$ at the horizon crossing during re-entry after the end of inflation and reheating. The usual curvature perturbations from inflation are expected to be nearly scale invariant and are constrained by cosmic microwave background (CMB) observations~\cite{Aghanim:2018eyx} at large scales. To allow formation of PBHs in this class of models, one must dramatically enhance the perturbations on some smaller scale, as determined by the PBH mass. This can be accomplished by introducing new features within the inflaton potential, usually at the expense of significant fine-tuning~\cite{GarciaBellido:1996qt,Kawasaki:2018daf,Germani:2018jgr}.  A hallmark of this class of models is that the PBH mass is close to the horizon mass at the time when large perturbations reenter the horizon.  Some production mechanisms suggested invoking metastability of the Standard Model vacuum~\cite{Espinosa:2017sgp}.  We will focus on a very different class of scenarios. 

Scalar fields that have large vacuum  expectation values (VEVs) in de Sitter universe (i.e.~during inflation) eventually relax to a minimum of the effective potential after the end of inflation. This process often leads to instabilities and subsequent fragmentation of the initially homogeneous scalar field condensate into solitonic lumps and free particles in plasma, which was studied in detail in connection with scalar fields predicted by  supersymmetry (SUSY)~\cite{Kusenko:1997si,Dine:2003ax}. The basic physics of such instabilities is similar to the origin of Jeans instability, which develops due to the attractive nature of gravity. In the case of a scalar field, the attractive self-interactions inside the condensate may be stronger than gravity, while the same interactions in empty space do not generate long-range forces since the scalar mediator is massive. Solitonic lump formation from the scalar field instabilities requires the scalar potential to be shallower than quadratic, ensuring that attractive field self-interactions are present. 

\begin{figure*}[htb]
\begin{minipage}[b]{0.5\textwidth}
\centering
\includegraphics[width=\textwidth]{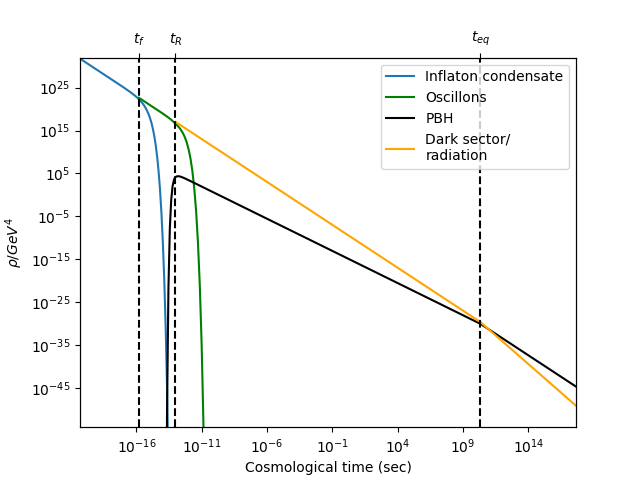}
\end{minipage}
\hspace{-1.5em}
\begin{minipage}[b]{0.5\textwidth}
\centering
\includegraphics[width=1.125\linewidth]{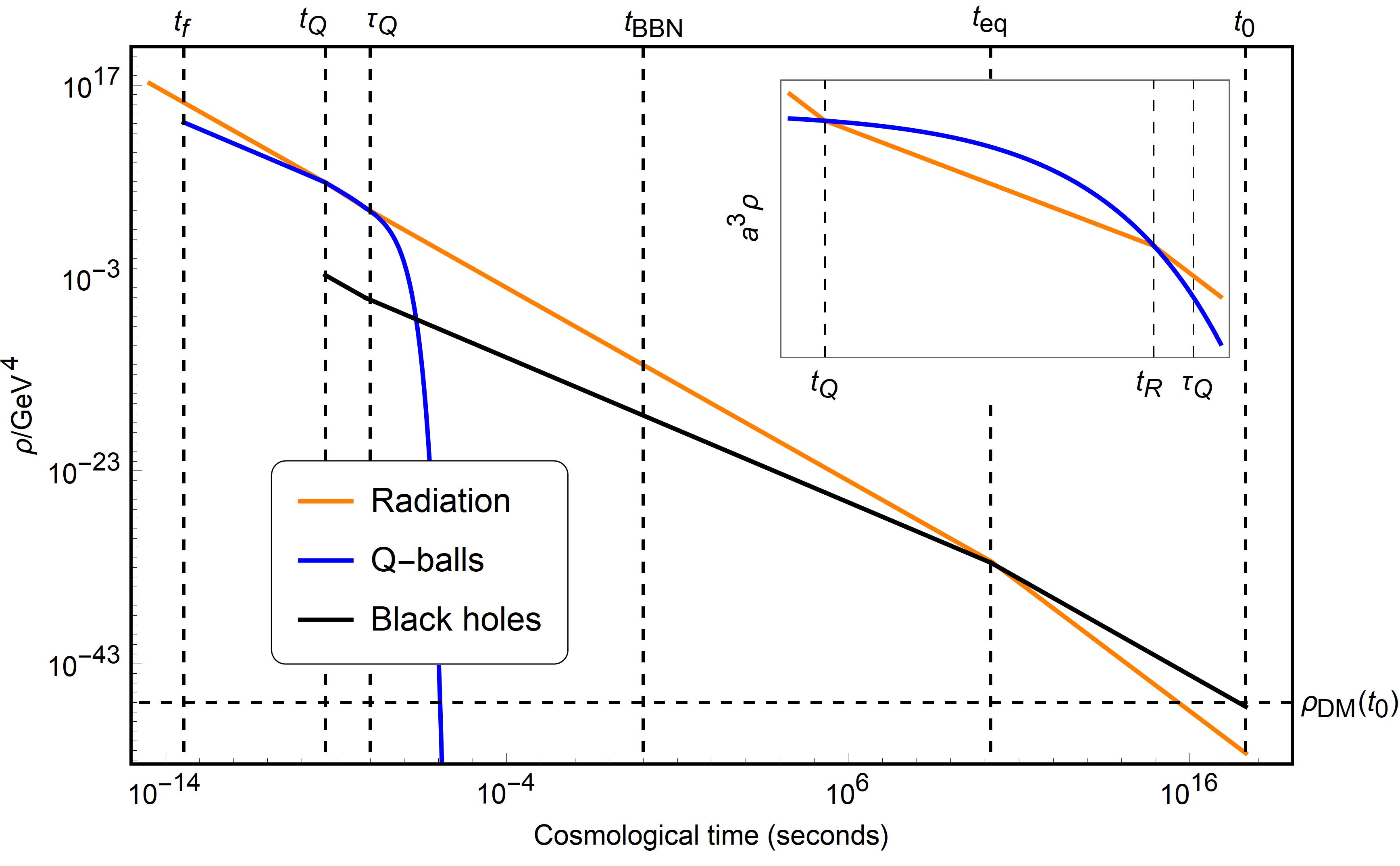}
\end{minipage}
		\caption{Two examples of a typical timeline. In the model from Ref.~\cite{Cotner:2018vug} (left panel),  oscillon-type scalar lumps dominate the energy density for a limited time creating a temporary matter-dominated era, during which  PBHs form. In the case of a spectator field~\cite{Cotner:2016cvr,Cotner:2017tir}, it can fragment into Q-balls during the radiation-dominated era (right panel). Since the gas of Q-balls has the energy density that scales as matter, it comes to dominate at time $t_Q$, creating a matter-dominated era that lasts until the Q-balls decay at time $\tau_Q$.  PBHs form during this matter-dominated era. 
	 }
		\label{fig:timeline}
\end{figure*}

The scalar condensate lumps formed in this process are sizable, and their distribution is stochastically generated. Hence, certain clusters of lumps have a sufficient overdensity to become building blocks for primordial black holes~\cite{Cotner:2016cvr,Cotner:2017tir,Cotner:2018vug}.
Since the source of perturbations generating PBHs on small scales is decoupled from the inflationary perturbations seeding large scale structure, it is not necessary to fine-tune the inflaton potential for the purpose of producing PBHs.~Unlike the standard PBH formation scenarios, PBH formation from fragmentation can occur after inflation either before or after the reheating has happened. While relativistic particles and other forms of radiation have energy density scaling of $\rho \propto a^{-4}$, with $a$ being the cosmological scale factor, the population of solitons scales as matter, i.e. $\rho \propto a^{-3}$. Thus, the matter in the form of the fragments can dominate the energy density at some later time.  Eventually, the scalar lumps decay, but the temporary matter-dominated era during which the density fluctuations are large, unrelated to the large-scale structure, and unconstrained by the CMB allows for production of PBH.  The timeline common to this class of models is illustrated in Fig.~\ref{fig:timeline}, where the matter in the form of scalar lumps (oscillons~\cite{Cotner:2018vug} or Q-balls~\cite{Cotner:2016cvr,Cotner:2017tir}) dominates the energy density for a short time, during which PBH are produced. 

Since PBHs form during a short matter-dominated phase from matter-like lumps, the associated angular momentum leads to PBHs with a higher spin (described by spin parameter $a_s$) than expected from the radiation-era production scenarios. We provide a comparative overview of the two types of PBH production scenarios in Table~\ref{tab:pbhcomp}.

The perturbations resulting from scalar field fragmentation are a form of isocurvature perturbations. However, unlike the perturbations generated by the inflaton or in models with a spectator curvaton field~(e.g.~\cite{Ando:2017veq}), these  perturbations are produced within the horizon {\it after the end of inflation}. Therefore, they do not affect the large-scale structure and any growth of perturbations during the temporary matter-dominated era associated with the scalar fragmentation takes place at the scales too small to have any effect on the CMB. 
 
\begin{table}[tb]
    \centering
    \addtolength{\leftskip} {-1cm}
    \addtolength{\rightskip} {-1cm}
\begin{tabular}{|l|c|c| } \hline
  & \multicolumn{2}{c|}{ \textbf{PBH Production Scenario}} \\ 
 & \underline{Inflationary Perturbations}	& \underline{Field Fragmentation} \\ 
 & \textit{(common mechanism)}	& \textit{(our mechanism)}  \\
\hline 
\hline
Source and type of large & inflaton fluctuations, & inflaton fluctuations,   \\
(CMB-scale) perturbations & curvature & curvature   \\
\hline 
Source and type of small & inflaton fluctuations, & stochastic field fragmentation,   \\
(PBH-scale) perturbations & curvature & isocurvature (fragment-lumps)   \\
\hline
  PBH source field  & inflaton & inflaton or spectator field  \\
 \hline
   \multirow{5}{*}{Required potential condition}  &  \multirow{5}{*}{inflaton potential fine tuning} & no new restrictions on inflaton \\& &  potential, scalar field potential     \\
&  &   shallower than quadratic     \\
&   &     (attractive self-interactions) \\
\hline
PBH formation era $(t_{\rm PBH})$ & $t_{\rm BBN} \gtrsim t_{\rm PBH} \gtrsim t_{\rm reh}$,   &  $t_{\rm BBN} \gtrsim t_{\rm PBH} \gtrsim t_{\rm inf}$,  \\
and type    & after reheating,  &  before or after reheating,  \\
 & radiation-dominated era & temporary matter-dominated era \\
\hline
PBH size $(r_{\rm BH})$ vs. horizon $(r_{\rm H})$ & \multirow{2}{*}{$ r_{\rm BH} \sim r_{\rm H} \sim H^{-1} $} &    \multirow{2}{*}{$ r_{\rm BH} \ll r_{\rm H} \sim H^{-1} $} \\
 at formation &    &      \\
\hline
PBH spin $(a_s)$  & $a_s \sim 0$ &  $a_s \sim \mathcal{O}(1)$ possible \\
 \hline
\end{tabular}
\caption{Comparative summary of standard PBH formation mechanism from inflationary perturbations and PBH production from scalar field fragmentation mechanism.}
\label{tab:pbhcomp}
\end{table}

In this work we describe in generality the mechanism of PBH production based on scalar field fragmentation into solitonic lumps. The mechanism can be realized within a broad class of particle physics models. The scalar field that undergoes fragmentation can be either a spectator or the inflaton,  and can be associated with a broad range of inflationary models.  Previous studies focused on the special cases of a complex scalar field fragmenting into Q-balls~\cite{Cotner:2016cvr,Cotner:2017tir} and a real scalar field fragmenting into oscillons~\cite{Cotner:2018vug}. We generalize these results and provide a universal analytic description for PBH formation from solitonic energy lumps with an arbitrary number of conserved discrete or continuous quantities, such as field charge. In addition, we elucidate the relation between the model parameters and the sizes of the resulting black holes.  In particular, we show that supersymmetric flat directions produce PBH in the range of masses allowed for dark matter. 

The paper is organized as follows. The fragmentation of a scalar field is described in Section~\ref{sec:frag}. We then present a general framework for PBH production in Section~\ref{sec:pbhform}. We summarize in Section~\ref{sec:sum}.

\section{Formation and fragmentation of a scalar condensate}
\label{sec:frag}

Assuming inflation took place in the early universe, followed by reheating and Friedmann--Robertson--Walker expansion, scalar fields with relatively shallow potentials (or relatively small masses) that are present in the theory undergo a non-trivial time evolution.~Such fields tend to develop non-zero vacuum expectation values during inflation, followed by a coherent evolution towards the minimum of the effective potential when inflation is over.  The latter is a process in which the field may or may not evolve in a spatially homogeneous manner.  Depending on the scalar potential, an initially homogeneous field may either remain homogeneous or fragment into lumps. The instability giving rise to the fragmentation occurs when the scalar self-interaction is attractive and some other auxiliary conditions are met.  Overall, the fragmentation is a fairly generic phenomenon and is expected to occur in a large variety of models, including supersymmetric theories. 

\subsection{Spectator fields in de Sitter universe}

We assume inflation in the early universe driven by the inflaton field $\Phi_I$ with the potential $U_I(\Phi_I)$, which leads to density perturbations consistent with observations.  We will not make any additional assumptions about the details of the inflationary model. We shall focus on a {\em different} (spectator) scalar field $\phi$ with a potential $U(\phi$) such that it makes a negligible contribution to the total energy density of the universe during inflation and subsequent reheating, that is
\begin{equation}
U_{\rm total}= U_I(\Phi_I)+U(\phi) \approx U_I(\Phi_I)~.
\end{equation}
If the scalar potential of the spectator field $U(\phi)$ is relatively shallow, the field develops a non-zero average  expectation value when averaged over super-horizon scales~\cite{Bunch:1978yq,Linde:1982uu,Lee:1987qc,Starobinsky:1994bd,Dine:2003ax}.  In de~Sitter space, there is a non-zero probability of tunneling from a lower to a higher value of the effective potential~\cite{Lee:1987qc}.  Therefore, a field with an initial zero value at the minimum of the effective potential can develop a non-zero value by a quantum tunneling event.  The subsequent relaxation of such a field to the potential minimum can occur by a coherent motion. However, if the potential is not very steep, the time-scale associated with this motion is much longer than the rate of tunneling events, implying that the resulting average VEV of $\phi$ remains non-zero.  

In the simplest case of a zero potential, $U(\phi)=0$, the field undergoes a random walk forming a Bunch-Davies vacuum: $\langle |\phi|\rangle \propto t^{1/2}$~\cite{Bunch:1978yq}.  
For non-zero potentials the expectation value is such that every scalar degree of freedom carries energy density determined by the Hubble parameter or the ``Gibbons-Hawking temperature"~\cite{Bunch:1978yq,Linde:1982uu,Lee:1987qc,Starobinsky:1994bd,Dine:2003ax}:

\begin{equation}
U(\langle |\phi |\rangle) \sim H_I^4 ~, 
\label{eq:initial_VEV}
\end{equation}
where  
\begin{equation}
 H_I =  \sqrt{\frac{U_{\rm total}}{3 M_{pl}^2}} \approx \sqrt{\frac{U_I(\Phi_I)}{3 M_{pl}^2}}
 \end{equation}
 and $M_{pl}$ is the reduced Planck mass. This average value of $\langle \phi \rangle$ serves as the initial condition for post-inflationary physics. As seen from Eq.~\eqref{eq:initial_VEV}, a shallower potential leads to a larger initial VEV.  The field maintains this value until the Hubble parameter $H$ decreases and the time-scale for the coherent motion becomes shorter than the Hubble time.  

\subsection{Scalar field instabilities: heuristic arguments}

We will now outline several heuristic arguments to provide an intuitive understanding of the general features of the scalar evolution in the presence of self-interactions.

A scalar field with a non-zero VEV at the end of inflation must relax to the minimum of its effective potential.  
It is insightful to consider the connection between the scalar field instabilities and the gravitational instability of self-gravitating gas. Namely, the scalar instability is similar to gravitational Jeans instability that is studied in astrophysics and cosmology in connection with the collapse of gas clouds and subsequent star formation. The attractive self-interaction of gravity makes the homogeneous distribution of particles unstable with respect to formation of dense clumps. It is well known that the associated instability occurs on length scales larger than the Jeans length ($\lambda_J \sim 1/k_J$), or for the wave numbers 
\begin{equation}
0<k<k_{\rm max}=k_{\rm J}= 2 (\pi G \rho)^{1/2}/c_s ~,
\end{equation}
where $\rho$ is the energy density, $c_s$ is the speed of sound and $G$ is the gravitational constant.

Analogously, a homogeneous scalar condensate can be unstable with respect to fragmentation into lumps in the presence of attractive self-interactions. In the case of a scalar field, self-interaction determined by the scalar potential can either stymie or enhance the instability.  A repulsive self-interaction can act to counter gravity and keep the distribution homogeneous, while an attractive interaction can provide an attractive force that is stronger and more important than gravity, leading to fragmentation.  The relevant forces exist inside the condensate and are mediated by the exchange of a scalar field.  If the scalar is massive, there may not be any long-range forces in empty space, while the presence of long-range or sufficiently long-range modes inside the condensate could play a more important role than gravity when it comes to stability and fragmentation.  
 
Let us consider a complex scalar field $\phi(x)$ with a Lagrangian
\begin{equation}
{\cal L}=|\partial_\mu \phi|^2 - U(\phi) ~.
\end{equation}
A spatially homogeneous solution of the equations of motion can be characterized by density $\rho$ and pressure $p$, which represent the diagonal components $T_0^0$ and $T_i^i$ (no sum) of the energy-momentum tensor, respectively:
\begin{eqnarray}
    \rho= |\partial_\mu \phi|^2+ U(\phi) \\
    p= |\partial_\mu \phi|^2 - U(\phi)
\end{eqnarray}
For a potential $U(\phi)= {\rm const} |\phi|^n $, the pressure is related to energy density by a simple relation~\cite{Turner:1983he}   
\begin{equation}
    p = \Big( \frac{n-2}{n+2}\Big)\rho 
\end{equation}
that has a straightforward physics interpretation. A non-interacting scalar field has potential $U(\phi)\propto \phi^2 $ and $n = 2$, so the scalar condensate behaves as a gas of particles at rest with zero pressure.  A potential that is effectively rising faster than the second power of the field  (i.e. $n > 2$)  represents a repulsive self-interaction, which generates a positive pressure.  Finally, a potential that rises effectively slower than the second power of the field  (i.e. $n < 2$)  describes an attractive interaction, resulting in negative pressure. 

In the presence of gravity, a classical pressureless scalar field is subject to gravitational instability on all scales. The reason is that as long as $p=0$  the speed of sound is zero, resulting in $k_J=\infty$.  However, due to Heisenberg uncertainty principle,  quantum effects result in a finite Jeans mass for a non-interacting scalar field~\cite{Khlopov:1985jw,Bianchi:1990mha}, hence
$k_J=(32 \pi^2 G U'' \rho)^{1/4}$.   For a repulsive self-interaction, for example in case of  $U(\phi)=m^2 |\phi|^2+\lambda |\phi|^4$ with $\lambda>0$, the repulsive forces can counter the gravitational attractive force and either reduce the range of the instability or eliminate it completely~\cite{Khlopov:1985jw}.  

An attractive self-interaction corresponds to a potential that effectively grows slower than the second power of the field.  Such interactions result in negative pressure, which increases the value of $k_J$, opening a wider spectral window for the instability.   When the self-interaction is stronger than gravity inside the scalar condensate, it is the scalar interaction that plays the leading role in driving the instability.  

The above heuristic arguments  highlight why the spatially homogeneous solution can be unstable in presence of attractive self-interactions. However, as we will discuss below, the shape of the potential, its symmetries, and the nature of the homogeneous solution play a central role~\cite{Kusenko:1997si}.  
\subsection{Which instabilities can lead to black hole formation?}

When considering PBH formation, one needs to know the end result of the instability.  An initially homogeneous condensate can fragment into stable or relatively stable Q-balls~\cite{Kusenko:1997si} or oscillons~~\cite{Bogolyubsky:1976nx,Gleiser:1993pt,Copeland:1995fq,Fodor:2006zs,Kasuya:2002zs,Honda:2001xg,Amin:2011hj,Hong:2017ooe}, which scale as cosmological matter and can lead to formation of PBHs.  Alternatively, the result of the instability may be a highly inhomogeneous state in which there are no stable high-density lumps and, in fact, the higher density regions may exhibit a higher rate of the condensate decay.  

\begin{figure*}[htb]
\centering
\includegraphics[width=0.475\textwidth]{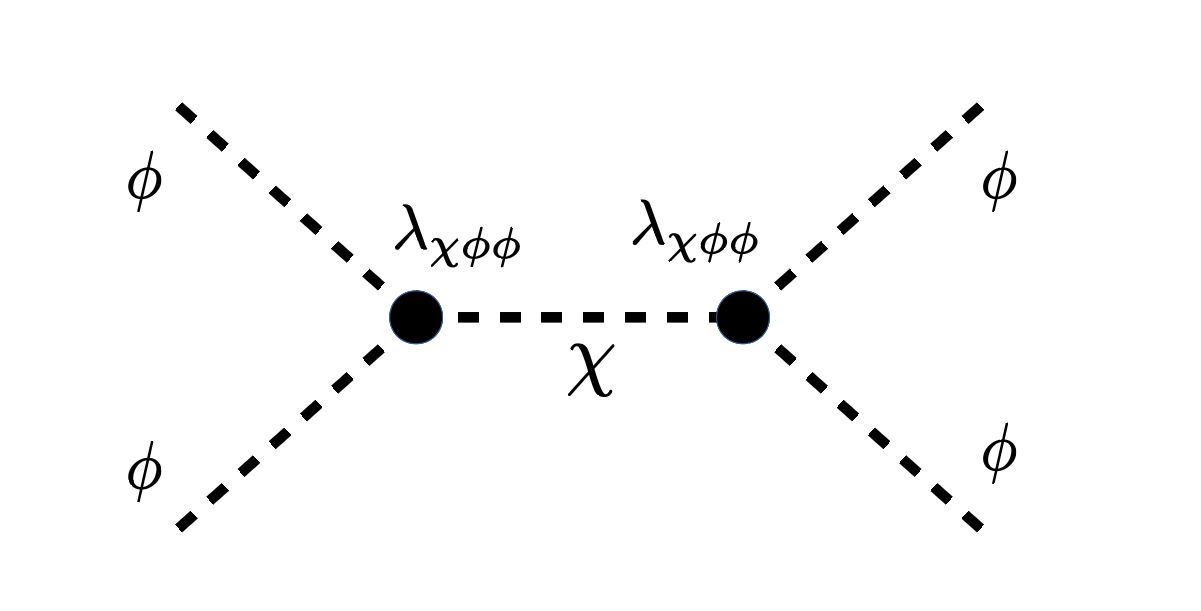}
\centering
\includegraphics[width=0.475\textwidth]{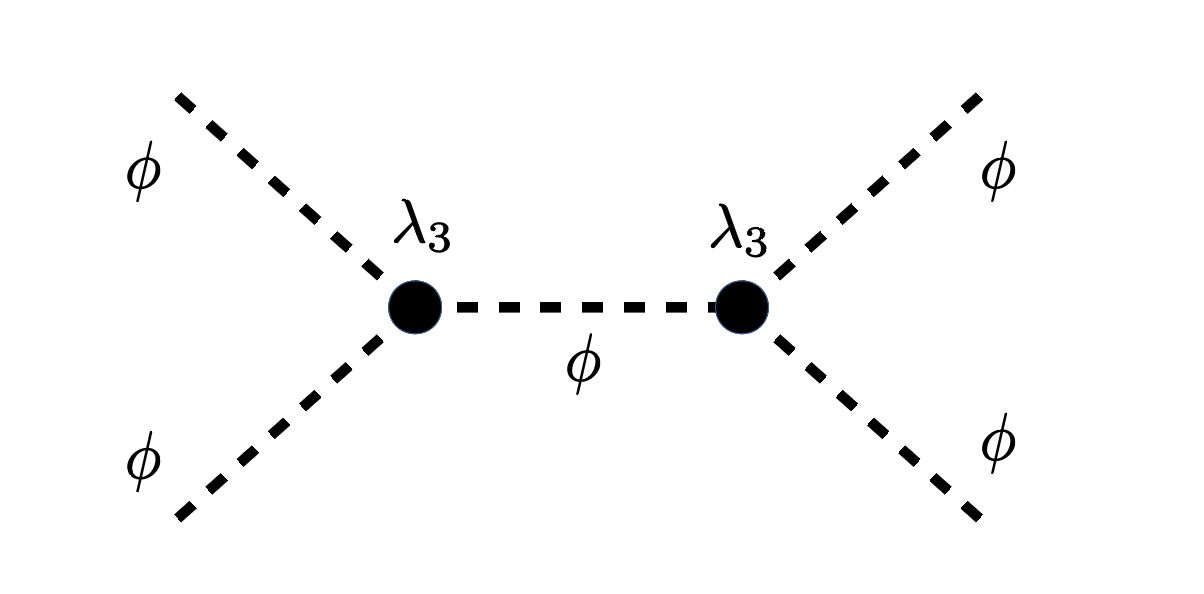}
		\caption{Attractive interactions mediated by a massive scalar exchange via a tri-linear vertex  $\lambda_{\chi\phi\phi} \chi \phi^\dag \phi $ conserve the particle number and can lead to formation of Q-balls~\cite{Coleman:1985ki} in theories with multiple scalar fields~\cite{Kusenko:1997zq}.  Coupling  $\lambda_3 \phi^3$ also produces attractive interactions, but, in general, it does not preserve the particle number. Nevertheless, such interactions can lead to localized long-lived pseudo-solitons, such as oscillons~\cite{Bogolyubsky:1976nx,Gleiser:1993pt,Copeland:1995fq,Fodor:2006zs,Kasuya:2002zs,Honda:2001xg}, whose stability can be traced to an adiabatic invariant~\cite{Kasuya:2002zs} or to an approximate symmetry in a theory that has the same non-relativistic limit~\cite{Mukaida:2014oza,Ibe:2019vyo}.}
		\label{fig:selfint}
\end{figure*}

At the level of Feynman diagrams, one can think of an attractive scalar self-interaction as an exchange of a scalar mediator via a tri-linear vertex (see Fig.~\ref{fig:selfint}). A cubic interaction of the form $U(\phi) \supset \lambda_3 \phi^3 $ is attractive, and it can cause an instability. However, the same interaction is responsible for particle-number changing processes $ \phi \phi \rightarrow \phi$. Such reactions can take place in the condensate at the rate that is unsuppressed, as long as the phase space occupied by the quanta in the condensate allows the $2\rightarrow 1$ processes to proceed unimpeded.  While a larger value of $\lambda_3$ is beneficial for instability and fragmentation, the same coupling also controls the decay of the scalar lumps, which can be detrimental for the PBH formation. This obstacle to PBH formation can be circumvented if the particle non-conserving processes are suppressed either by a symmetry that enforces the particle number conservation (the case of Q-balls), or if the scalar condensate fragments into solitons in which the quanta occupy a peculiar phase space suppressing the $2\rightarrow 1$ processes (the case of oscillons).  

{\it Q-balls}. If a theory possesses a U(1) symmetry, there is a stable Q-ball solution in a scalar potential that allows an attractive interaction~\cite{Coleman:1985ki}.  Since odd powers do not respect the global U(1) symmetry $\phi \rightarrow e^{i\theta}\phi$, it was thought that renormalizable theories in 3+1 dimensions do not allow for Q-balls.  However, such solitons are allowed in theories that contain more than one field, if some of the fields have a zero U(1) charge~\cite{Kusenko:1997zq}. The requisite attractive interaction can be given, for example, by an effective vertex $U(\phi) \supset \lambda_{\chi\phi\phi} \chi \phi^\dag \phi $ that respects the U(1) symmetry $\phi \rightarrow e^{i\theta}\phi, \chi \rightarrow \chi$ and, therefore, allows for a particle number conservation.   Such self-interaction can, indeed, lead to fragmentation of a scalar condensate into Q-balls, which can be stable or long-lived and can merge into PBH. 

{\it Oscillons}. The longevity of oscillons is supported by a dynamical invariant that exists for a class of potentials for a real scalar field~\cite{Kasuya:2002zs}.  These solitons are related to Q-balls in that an oscillon can be described as a Q-ball in a related theory with a complex scalar field and an approximate U(1) symmetry in the non-relativistic limit~\cite{Mukaida:2014oza,Ibe:2019vyo}.  Therefore, fragmentation of a real scalar field into oscillons bears some similarities to the complex field fragmentation into Q-balls discussed below.  

\subsection{Instability of scalar field with a U(1) symmetry}
	
We will now discuss some general features of the relevant instabilities, which can be studied by means of Floquet analysis,  following Ref.~\cite{Kusenko:1997si}. Let us consider a complex scalar field $\phi(x,t)$ that does not dominate the energy density.~Generalizations to a real scalar field and the inflaton are analogous. To explore the scalar potential $U(\phi)$ with a minimum at $\phi=0$ and that respects a U(1) symmetry corresponding to a conserved particle number, it is helpful to reparametrize the field as a real scalar field with a time-dependent complex phase: $\phi=\Phi(x,t) e^{ i\Theta(x,t) }$. The spatially homogeneous solution to the equations of motion is described by
\begin{equation}
    \phi(x,t)=\phi(t) \equiv \Phi(t) e^{i \Theta(t)} ~. 
\end{equation}
The field starts at some initial value, determined by high-scale physics and inflation, and then relaxes to the potential minimum.  Let us examine the stability of such a solution with respect to small perturbations. The classical equations of motion in the spherically symmetric metric $ds^2=dt^2-a(t)^2 dr^2$ are
\begin{eqnarray}
	\left \{ \begin{array}{lll}
	\ddot \Theta +3 H \dot \Theta - \dfrac{1}{a^2(t)} \Delta \Theta + \Big(\dfrac{2
		\dot  \Phi}{\Phi} \Big)
	\dot \Theta - \dfrac{2}{a^2(t) \Phi} (\partial_i \Theta)(\partial^i \Phi) & = & 0~,   
	\\
	\ddot \Phi + 3 H\dot \Phi - \dfrac{1}{a^2(t)} \Delta \Phi - 
	\dot \Theta^2 \Phi + \dfrac{1}{a^2(t)} (\partial_i \Theta)^2
	\Phi + \Big(\dfrac{\partial U}{\partial \Phi}\Big) & = & 0~,
	\end{array} 
	\right. 
	\label{eqnmtn1}
	\end{eqnarray}
where dots denote the time derivatives, and the space coordinates are
	labeled by the Latin indices that run from 1 to 3. The Hubble parameter
	$H=(\dot a/a)$, where $a(t)$ is the scale factor, is proportional to $t^{2/3}$
	or $t^{1/2}$ for the matter or radiation-dominated universe,
	respectively.
 
    To study the stability of a homogeneous solution, we add   small perturbations $\delta \Phi, \delta \Theta \propto e^{S(t) - i \vec{k}\vec{x}} $ to the system and look for
	growing modes, $({\rm Re} \, \alpha>0)$, where $\alpha = dS/dt $. The wave number $k$ in the comoving frame is
	red-shifted with respect to the physical wave number $\tilde{k}=k/a(t)$ in the expanding background. 
	
	From the equations of motion, one can derive the equations for small perturbations and linearize them by neglecting any terms quadratic in $\delta \Phi, \delta \Theta$:
\begin{eqnarray}
	\left \{ 
	\begin{array}{ll}
	&\ddot{\delta \Theta} + 3 H \dot{(\delta \Theta)}
	- \dfrac{1}{a^2(t)} \Delta (\delta \Theta) +\Big(\dfrac{2 \dot
		\Phi}{\Phi}\Big) \dot{(\delta \Theta)}+ \Big(\dfrac{2 \dot \Theta }{\Phi}\Big)
	\dot{(\delta \Phi)}
	- \Big(\dfrac{2\dot \Phi \dot \Theta }{\Phi^2}\Big) \delta \Phi
	  ~=~   0~, \label{eqndelta1} \\
	&\ddot{\delta \Phi} + 3 H \dot{(\delta \Phi)}
	- \dfrac{1}{a^2(t)} \Delta (\delta \Phi)
	-2 \Phi \dot \Theta \dot{(\delta \Theta)}+ U'' \delta \Phi - \dot \Theta^2
	\delta \Phi   ~=~   0~.
	\end{array} 
	\right. 
	\label{eqnmtn1}
	\end{eqnarray}
	The above implies the following dispersion relation, where we define  $\omega \equiv \dot \Theta $ :
	\begin{eqnarray}
	\left [
	\alpha^2+ 3H\alpha +\frac{k^2}{a^2} + \Big(\frac{2 \dot \Phi}{\Phi}\Big) \alpha \right
	]
	\left [ \alpha^2+ 3H\alpha +\frac{k^2}{a^2} +U''(\Phi)- \omega^2
	\right ]+
	4 \omega^2 \left [ \alpha - \frac{\dot \Phi}{\Phi} \right ] \alpha = 0~.    
	\label{eq:disprel}
	\end{eqnarray}
We want to identify the range of modes $k$ for which the Lyapunov exponents $\alpha(k)$ are positive, leading to an exponentially growing instability.  Examining the zeros of $\alpha$, Eq.~\eqref{eq:disprel} becomes
\begin{eqnarray}
\frac{k^2}{a^2} 
	\left [\frac{k^2}{a^2} + U''(\Phi) - \omega^2 
	\right ]	= 0~,  \ \   {\rm with }\ \alpha = 0.
	\label{dr2}
\end{eqnarray}
If the potential has a negative second derivative $U''<0$ or if the homogeneous solution has a large enough $\omega > U''$, then the following inequality is satisfied 
\begin{equation}
\tcboxmath{
U''(\Phi) -    {\omega}^2  < 0~,}
    \label{eq:instcond}
\end{equation}
leading to two non-negative values of $k$ that solve Eq.~\eqref{eq:disprel}, namely $k=0$ and $k=k_{\rm max}\equiv a \sqrt{\omega^2-U''(\Phi) } $. Therefore, the function $\alpha (k)$ has only two roots and $\alpha (k)$ must have a constant sign for $0<k<k_{\max}$.  By examining the sign of $\alpha $ in the limit $k\rightarrow 0^+$, one can see that $\alpha(k)>0$ over the entire range $0<k<k_{\max}$.  
This implies there is a band of unstable modes given by 
\begin{equation}
 \tcboxmath{{\rm \underline{unstable\ modes}:}\ \ \alpha(k)>0, \ \ {\rm for} \  \ 0<k<k_{\max}  = \sqrt{\omega^2-U''(\Phi) }.}
\end{equation}
If the condition of Eq.~\eqref{eq:instcond} is not satisfied, then all the modes are stable. In the case of a real field or a complex field with no time-dependent phase, when $\omega = \dot{\Omega}=0$, the instability discussed above reduces to the well known tachyonic instability: 
	\begin{equation}
	U''( \Phi )  < 0~.
	\label{eq:condtach}
	\end{equation}
	
We conclude that a broad class of potentials leads to an instability and fragmentation into Q-balls.  These potentials include the tachyonic potentials (e.g. the logarithmically growing potentials in gauge-mediated supersymmetry models) as well as potentials with a positive second derivative, which can also lead to a fragmentation into Q-balls if the condensate has a sufficiently high  $\dot{\Theta}$, implying a  sufficiently high particle number density $n= \dot{\Theta} \Phi^2$. 

\subsection{Solitonic fragments}
\label{ssec:qballs}

The instability described above in the linear regime can develop and become non-linear (i.e. $\delta \phi \gtrsim \overline{\phi})$ if it grows fast enough in comparison with the expansion rate of the universe.  The end result of this evolution can be a spatially inhomogeneous state consisting of Q-balls~\cite{Coleman:1985ki} or oscillons~\cite{Bogolyubsky:1976nx,Gleiser:1993pt,Copeland:1995fq,Fodor:2006zs,Kasuya:2002zs,Honda:2001xg}.  

For a complex scalar field with a potential that respects a global U(1) symmetry, the end result of the evolution is a distribution of non-topological solitons, Q-balls, which minimize the energy for a fixed particle number~\cite{Coleman:1985ki}.  These can be thought of as coherent bound states of scalar particles.
	
All supersymmetric extensions of the Standard Model contain Q-balls in their spectrum, made of scalar baryons and leptons predicted by supersymmetry~\cite{Kusenko:1997zq}. They can be entirely stable if the scalar field carrying the baryon number extends along a flat direction of the scalar potential~\cite{Dvali:1997qv,Kusenko:1997si}. As already noted, the size of the Q-balls or other lumps resulting from fragmentation is determined by the fastest growing mode that becomes non-linear the earliest, denoted $k_{\rm nl}$.  Based on numerical simulations\footnote{These numerical simulations have been performed for  scalar fields that arise in supersymmetric models, and the fragmentation was considered in the context of Affleck-Dine baryogenesis and dark matter production.  While the flat directions associated with supersymmetry are well suited for PBH production, both analytical and numerical results are applicable to a broad range of scalar fields. The simulations have been performed in both 1+1 as well as in full 3+1 dimensions and taking the expansion of the Universe into account. }~\cite{Kasuya:2000wx,Multamaki:2002hv, Kusenko:1997si}, the size $R_Q$ of a typical lump constitutes a few percent of the horizon size:   
	\begin{equation}
	R_Q \sim k^{-1}_{\rm nl} \sim f_Q H^{-1}~,
	\end{equation}
	where $f_Q\sim 10^{-2}-10^{-1}$. Since the resulting soliton number density at fragmentation time is approximately
	\begin{equation} \label{eq:nsol}
    \overline{n} \sim \bigg(\dfrac{k_{nl}}{2 \pi}\bigg)^3~,
\end{equation}
    we expect the total number of fragments filling the volume to be in the range $\sim  10 - 10^6$. 
	The matter component composed of scalar lumps generated at the time of fragmentation can later come to dominate the energy density. The nature of the fragments depends on the potential.
	
Oscillons, pseudo-solitonic bound field configurations~\cite{Bogolyubsky:1976nx,Gleiser:1993pt,Copeland:1995fq,Fodor:2006zs,Kasuya:2002zs,Honda:2001xg}, can form as a result of a real scalar field fragmentation. The oscillons arise in many well motivated theories with scalar fields, such as models of inflation~\cite{Amin:2011hj,Hong:2017ooe}, axions~\cite{Kolb:1993hw} or moduli~\cite{Antusch:2017flz}.  The oscillons are localized and metastable. The stability of an oscillon is due to  an  approximate adiabatic invariant~\cite{Kawasaki:2015vga,Kasuya:2002zs}, which can be described as an approximately conserved particle number in a non-relativistic limit of a related theory with a complex scalar field~\cite{Mukaida:2014oza,Ibe:2019vyo}.  Early universe oscillons have been recently studied in connection with primordial gravity waves \cite{Antusch:2016con} as well as baryogenesis~\cite{Lozanov:2014zfa}.
	
The solitonic lumps we are considering are significantly smaller than the size of the horizon at fragmentation time. This allows us to treat them effectively as point particles and renders the effects of gravitational self-interactions for each individual fragment formation as negligible. Hence, gravitational effects become only relevant for a cluster of lumps and the subsequent collapse to a black hole, while the dynamics of formation of individual lumps (e.g. instabilities and fragmentation) are governed solely by the field's self-interactions not associated with gravity (potential shallower than quadratic). In the presence of gravitational field interactions and curved space-time non-linear field theories can also admit massive macroscopic gravitational soliton configurations, which can be collectively referred to as ``soliton stars'' (for review see e.g.~Ref.~\cite{Lee:1991ax,Schunck:2003kk}). They have been studied extensively in the literature under various names and in different specific realizations, including boson and fermion stars~(e.g. \cite{Ruffini:1969qy,Lee:1991ax,Henriques:1989ez,Kaup:1968zz}, oscillatons~\cite{UrenaLopez:2001tw}, axion stars and the related mini-clusters~(e.g. \cite{Kolb:1993zz,Braaten:2018nag}). Collapse to a black hole as well as general stability of individual soliton stars has been extensively explored (e.g. Ref.~\cite{Helfer:2016ljl,Braaten:2015eeu}). Due to their size, we do not expect a significant amount of such soliton stars to be formed within a single horizon. Hence, in such case, it will be a non-trivial task to achieve large density perturbations due to a cluster of soliton stars within some volume in order to produce PBHs via the mechanism described in this work. We shall thus focus solely on non-gravitationally coupled soliton configurations.
	
\section{Primordial black hole formation from lumps of scalar field}
	\label{sec:pbhform}
	
	\subsection{General framework}

	We are interested in finding the expected PBH energy density $\braket{\rho_\text{PBH}}$ as a function of PBH mass $M$, i.e. the PBH~\textit{ mass function}, resulting from gravitational collapse of a collection of soliton fragments. Solitons residing in a group (a cluster) whose local density $\rho$ is larger than the soliton background density $\langle \rho \rangle$,  when the fractional overdensity fluctuation is
	\begin{equation} \label{eq:denfluc}
	    \delta = \dfrac{\rho-\braket{\rho}}{\braket{\rho}} > 0~,
	\end{equation}
tends to infall under the action of gravity and could form a black hole.
	
	If some fraction $B(M,V)$ of soliton clusters of mass $M$ residing within an initial volume $V$ will eventually collapse to black holes, then the expected average energy density of resulting BHs is given by
	\begin{equation}
	\braket{\rho_\text{PBH}}_V = \braket{B\rho_{\rm M}}_V = \braket{B M/V}_V = \dfrac{1}{V} \int dM\, P(M|V)\, B(M,V) M ~,
	\end{equation}
	where $\rho_{\rm M} = M/V$ denotes the cluster energy density and $P(M|V)$ is the probability of finding a cluster of mass $M$ within a given volume $V$.
This is the contribution due to a single volume scale as well as corresponding BHs of a particular size. To get the total average energy density, we sum over all volume scales and effectively over PBHs of all sizes, which would contribute to the process. The total resulting PBH energy density is
	\begin{gather} \label{eq:PBH_density}
	\braket{\rho_\text{PBH}} = \int d(\ln V) \braket{B \rho_{\rm M}}_V = \int \frac{dV}{V^2} \int dM\, P(M|V)\,  B(M,V) M~.
	\end{gather}
	
The above treatment allows to capture the behavior of soliton clusters with an arbitrary mass-energy distribution $P(M|V)$ and further input assumptions are required to make progress. In a typical situation the soliton cluster mass distribution is related to the number of individual solitons composing the cluster and whose number distribution is specified. Then, $P(M|V)$ can be decomposed as
\begin{equation} \label{eq:ndec}
P(M|V) = \sum_N P(M|N) P(N|V)~,
\end{equation}
where $P(M|N)$ is a distribution of soliton cluster mass $M$ given a soliton number $N$. Here, $P(N|V)$ is the number distribution of individual solitons within $V$, which we will assume to be Poisson. The decomposition of Eq.~\eqref{eq:ndec} is employed to treat PBH formation from oscillons.

In principle, other quantities aside the individual soliton number can have an effect on the mass-energy distribution of the soliton clusters. In the case of Q-balls, the additional dependency originates from the conserved Q-ball charge. This leads to a further conditional decomposition of $P(M|V)$ as 
\begin{align} \label{eq:qndec}
P(M|V) =& \sum_N \int dQ\, P(M|N,Q) P(N,Q|V) \notag\\ =& \sum_N \int dQ\, P(M|N,Q) P(Q|V) P(N|V)~.
\end{align}
We note that summation in $N$ accounts for soliton number being a discrete quantity, while the integration in $Q$ accounts for charge being continuous.
 
In the most general case, if the soliton cluster mass-energy depends on an arbitrary number of discrete quantities $Q_i$ forming a set $\{Q\}$ and continuous quantities $\xi_j$ forming a set $\{\xi\}$, then the respective mass distribution will be given by
	\begin{equation} \label{eq:gendec}
	    P(M|V) = \left( \prod_{i}  \sum_{Q_i} \right)  \left( \prod_{j} \int d\xi_j \right)\, P(M| \{Q\}, \{\xi\}) \, P(\{Q\}, \{\xi\}|V)~.
	\end{equation}
If the quantities $\{\xi\}$ and $\{Q\}$ are distributed independently, then $P(\{Q\}, \{\xi\}|V)$ can be separated further as in Eq.~\eqref{eq:qndec}, yielding
	\begin{equation} \label{eq:gendec2}
	    P(M|V) = \left( \prod_{i}  \sum_{Q_i} P(Q_i|V) \right)  \left( \prod_{j} \int d\xi_j P(\xi_j|V) \right) \, P(M| \{Q\}, \{\xi\}) ~.
	\end{equation}

The expressions above can be simplified if we assume that all solitons within a given volume $V$ are identical (denoted ``id''). That is, while the quantities in $\{Q\}$ and $\{\xi\}$ can vary independently, the resulting value of each quantity is the same for every soliton. Then,
	\begin{equation}
	    P_{\rm id} (M|\{Q\},\{\xi\}) = \delta(M - M_f (m|\{Q\}, \{\xi\})) P(m|\{Q\},\{\xi\}) ~,
	\end{equation}
	where $P(m|\{Q\},\{\xi\})$ is the probability of an individual soliton to have mass $m$ given $\{Q\}$ and $\{\xi\}$ and the soliton cluster mass is determined by a mass function of $m$, $M_f (m|\{Q\}, \{\xi\})$, which depends on the values of the quantities in $\{Q\}$ and $\{\xi\}$. In the simplest case of $N$ identical solitons of mass $m$ and no other dependencies, i.e. $\{Q\} = N$ and $\{ \xi \} = \emptyset$, one obtains
		\begin{equation}
	    P_{\rm id} (M|N) = \int dm \, \delta(M - m N) P(m) = \dfrac{P(M/N)}{N}~,
	\end{equation}
	where we have assumed that distribution of mass for an individual soliton is independent of the number of solitons present, i.e. $P(m|N) = P(m)$.
	
	In contrast, if the mass of solitons follows an identical distribution $P(m)$ but is determined for each individual soliton separately (i.e. solitons are independently and identically distributed, ``iid''), then the mass distribution of a soliton cluster composed of $N$ solitons follows
	\begin{align}
	    P_{\rm iid} (M|N) &= \int \frac{d\mu}{2\pi} e^{-i\mu M} \tilde{P}(\mu)
	    = \int \frac{d\mu}{2\pi} e^{-i\mu M} \int dM\, e^{i\mu M} P(M) \notag\\
	    &= \int \frac{d\mu}{2\pi} e^{-i\mu M} \int dM\, e^{i\mu M} \left(\prod_{i=1}^N \int dm_i\, P(m_i)\right) \delta\left(M - \sum_i m_i\right) \notag\\
	    &= \int \frac{d\mu}{2\pi} e^{-i\mu M} \left(\prod_{i=1}^N \int dm_i\, P(m_i) e^{i\mu m_i}\right) \notag\\
	    &= \int \frac{d\mu}{2\pi} e^{-i\mu M} \tilde{P}^N(\mu)~.
	\end{align}

The difference between the above two assumptions (id vs. iid) can be readily illustrated for the case of soliton masses being normally distributed, with a PDF for single soliton being
\begin{equation}
P_{\rm s}(m) = \dfrac{e^{-(m-m_0)^2/2\sigma^2}}{\sqrt{2\pi \sigma^2}}~,
\end{equation}
where $m_0$ is the mean mass of an individual soliton and $\sigma^2$ is the mass variance. The mass distribution of a soliton cluster with $N$ identical solitons then follows
	\begin{align}
	    P_{\rm id} (M|N) =&  \dfrac{P_{\rm s}(M/N)}{N} = \frac{\exp\left[-\frac{(M - Nm_0)^2}{2(N\sigma)^2}\right]}{\sqrt{2\pi (N\sigma)^2}}~, \\
	    P_{\rm iid} (M|N) =& \int \frac{d\mu}{2\pi} e^{-iM\mu} \tilde{P}_{\rm s}^N(\mu) = \frac{\exp\left[-\frac{(M - Nm_0)^2}{2(\sqrt{N}\sigma)^2}\right]}{\sqrt{2\pi (\sqrt{N}\sigma)^2}}~.
	\end{align}
The resulting expectation value $E$ and variance Var of a cluster mass, compared to single soliton case, are
\begin{align}
E(P_{\rm s})~=&~m_0 & \text{Var}(P_{\rm s})~=&~\sigma^2 \notag\\
E(P_{\rm id})~=&~N m_0 & \text{Var}(P_{\rm id})~=&~N^2 \sigma^2 \\
 E(P_{\rm iid})~=&~N m_0 & \text{Var}(P_{\rm iid})~=&~N \sigma^2~~.  \notag
\end{align}
While the id vs. iid mean cluster mass grows linearly in the number of solitons, the cluster mass spread is distinct for the two cases. This is just the standard probability theory result for a collection of $N$ iid variables with a Gaussian distribution. 
	
The field fragmentation process is stochastic, as supported by numerous simulation results (e.g.~\cite{Kusenko:1997si,Kasuya:2000wx,Multamaki:2002hv}). Hence, we can infer the soliton probability distribution within a given volume, $P(N|V)$, from statistical fluctuations alone. Given an average number
density $\overline{n}$ of uniformly distributed objects, the probability of finding $N$ objects within a volume $V$ follows the usual Poisson distribution
\begin{equation}
P(N|V) = \dfrac{(\overline{n}V)^N}{N!}e^{-\overline{n}V}~.
\end{equation}
The Poisson distribution originates from binomial distribution of $N$ independent random events in the large event
limit. This choice of $P(N|V)$ constitutes the most general description of fragmentation
assuming that events are uncorrelated and further it
allows for an analytic treatment. A deviation from this distribution would be highly model dependent and we thus do not consider this possibility further.

The resulting PBH mass function can be readily obtained from Eq.~\eqref{eq:PBH_density} as
\begin{equation}
\tcboxmath{\frac{d\braket{\rho_\text{PBH}}}{dM} = \int d(\ln V) \frac{d\braket{\rho_{\rm M}}_V}{dM} = \int \frac{dV}{V^2} P(M|V)\, B(M,V) M} 
\end{equation}
This gives the average energy density that goes into forming PBHs at the time of scalar field fragmentation. Once the black holes are formed and the universe becomes radiation dominated, the PBH mass distribution is relatively unchanged. To connect with observational constraints, the above PBH energy density must be redshifted into the future.

\subsection{Conditions for black hole formation from extended objects} \label{sec:NecessaryConditions}

Since solitons are extended objects, additional conditions are required to ensure that they can form BHs.
If fragmentation takes place at a time $t_f$, corresponding to a scale factor of $a_f$, solitons can come to matter-dominate the energy density at a later time $t_{\rm Q}$, corresponding to $a_{\rm Q}$, ending with decays resulting in a radiation-dominated era at $t_R$, when the scale factor is $a_{\rm R}$. In terms of redshift $z$ and time, this is
\begin{align}
\left(\dfrac{a_{\rm Q}}{a_f}\right)=&~z_{\rm fQ} +1 = \left(\frac{t_Q}{t_f}\right)^{1/2} = r_f^{1/2} \\
\left(\dfrac{a_{\rm R}}{a_{\rm Q}}\right) =&~z_{\rm QR} +1 = \left(\frac{t_R}{t_Q}\right)^{2/3} = r^{2/3}
\end{align}
where the factors $r_f=t_Q/t_f$ and $r=t_R/t_Q$ have been defined for future convenience. 

A collection of $N$, mass $M$ solitons, can form a BH with mass $M_{\rm BH} = N M_s$ at $a=a_{Q}$ if they are located within a Schwarzschild radius $R_{\rm BH}(N)=2N M_s/M_{\rm pl}^2$, with $M_{\rm pl}$ being the reduced Planck mass. The solitons must be sufficiently compact to fully fit inside the BH horizon and if one is to treat them particle-like. We require that each individual soliton exceeds the size of a BH with the same mass, but a collection of $N$ solitons can be fully enclosed in a volume of BH with mass $N M_s$ without solitons overlapping in space. Assuming all solitons are spherical with identical radius $R_s$, these restrictions translate to
\begin{equation} \label{eq:bhcond1}
\text{\underline{Black hole condition 1}:}~~~~~~~~~~ \dfrac{2 M_s}{M_{\rm pl}^2}< R_s< \dfrac{2 N^{2/3} M_s}{M_{\rm pl}^2}~.
\end{equation} 
In principle, the r.h.s. condition of Eq.~\eqref{eq:bhcond1} should also reflect the theoretical limit of how dense non-overlapping spheres can be packed into a given volume, which for 3-D Euclidean space constitutes $\sim75\%$ of the volume (see Kepler's theorem~\cite{Hales:2015}). We do not account for this restriction, which will not drastically alter our conclusions and that for our problem at hand might be further modified by GR-related corrections. Further, while the resulting solitons are identically randomly distributed and won't have the exactly same radius, from simulations (e.g.~\cite{Kusenko:1997si,Kasuya:2000wx,Multamaki:2002hv}) we expect that their size is strongly concentrated around the mean value, which we take to be the $R_s$ above.~We also note that, strictly, Eq.~\eqref{eq:bhcond1} is for resulting BHs with negligible rotation (i.e.~spin parameter $a_s\sim0$). Since PBHs from matter-dominated formation will typically have a non-negligible spin, the resulting BH radius $R_{\rm BH} = (1 + \sqrt{1-a_s}) G M_{\rm BH}$ can be up to a factor of 2 smaller. This effect will not drastically alter our analysis results and is not included.

Furthermore, particles separated by distances $\sim\,d$ can form a black hole if they are gravitationally bound and do not participate in the Hubble flow.  That is, the kinetic energy associated with the Hubble flow velocities $v\sim Hd $ is smaller than the  gravitational potential energy: 
\begin{equation} \label{eq:overlap}
\dfrac{H^2d^2}{2}<\frac{N M_s}{M_{\rm pl}^2 d}~. 
\end{equation}
Since the solitons are non-overlapping, the separation distances at the time of formation $d_f$ cannot be smaller than the soliton radius, $d_f>R_s$. The expansion of the universe stretches these distances at least by factor  $r_f^{1/2}$ until they can form a gravitationally bound system during the matter dominated era $d>R_s r_f^{1/2}$. Hence, condition of Eq.~\eqref{eq:overlap} implies
\begin{equation} \label{eq:bhcond2}
\text{\underline{Black hole condition 2}:}~~~~~~~~~~R_s<\left (\dfrac{2 N M_s}{M_{\rm pl}^2 H^2 r_f^{3/2}} \right )^{1/3}~,
\end{equation}
which for some parameter values can supersede the r.h.s. of condition in Eq.~\eqref{eq:bhcond1}.

The above conditions constrain the allowed mass-radius relations for solitons that are to compose BHs and for a given theoretical model they can be translated into restrictions on the allowed range of model parameters. For Q-balls, conditions of Eq.~\eqref{eq:bhcond1} and Eq.~\eqref{eq:bhcond2} can be directly translated into conditions for parameters of the potential through Eq.~\eqref{eq:flatpotential}, resulting in
\begin{equation} 	\label{eq:bhcondqball1} 
	1 < \frac{1}{2 Q^{\alpha-\beta}} \left(\frac{M_{\rm pl}}{\Lambda}\right)^2 < ~N^{2/3}
\end{equation}
and
\begin{equation} \label{eq:bhcondqball2} 
	    \Lambda  > \left(\frac{M_{\rm pl}^2 H^2 r_f^{3/2} Q^{3\beta-\alpha}}{2N}\right)^{1/4}~.
\end{equation}
Since for the scenario we will consider below  $3\beta-\alpha=0$, the above constraint will be independent of $Q$.
	
\subsection{Primordial black holes from supersymmetric Q-balls}
\label{sec:pbhqballs}

We illustrate the application of the general framework presented in Section~\ref{sec:pbhform} by outlining in detail PBH formation from supersymmetric Q-balls, following analysis of Ref.~\cite{Cotner:2016cvr,Cotner:2017tir} (see Ref.~\cite{Cotner:2018vug} for PBH formation from oscillons).
 
As mentioned, Q-balls appear in all SUSY extensions of the SM~\cite{Kusenko:1997zq}.
For instance, in gauge-mediated supersymmetry breaking (parametrized by the scale $M_{\rm SUSY}$), supersymmetric flat directions are lifted by a potential 
	\begin{equation}
	U(\phi)=M_{\rm SUSY}^4\log^2\left(1+\frac{|\phi|^2}{M_{\rm SUSY}}\right)~.
	\end{equation}
We note that this potential has a negative second derivative at large VEVs, so that the condition of Eq.~\eqref{eq:instcond} is automatically satisfied for any value of $\dot \Omega$. 
If the flat direction carries a conserved $U(1)$ number, such as a baryon or lepton number, then the mass of a Q-ball and its profile are determined by the amount of baryonic or leptonic charge $Q=B, L$ as
	\begin{align}
	M_Q =&~ M_{\rm SUSY} \ Q^{3/4} \\
	R_Q =&~ \dfrac{Q^{1/4}}{M_{\rm SUSY}} \notag
	\end{align}
	More generally, the mass and radius of a Q-ball are given by some mass scale $\Lambda$ and parameters $\alpha$ and $\beta$, with $0<\alpha$ and $\beta<1$, which depend on the shape of the scalar potential  
	\begin{gather}
	M_Q = \Lambda Q^\alpha \\
	R_Q = \dfrac{Q^\beta}{\Lambda}  \notag
	\label{eq:flatpotential}
	\end{gather}
	While Q-balls are stable with respect to the decay into the original scalar field particles, they could decay, for example, into lighter fermions~\cite{Cohen:1986ct,Kawasaki:2012gk,Enqvist:1998xd} or if the associated $\mathrm{U}(1)$ symmetry is broken by the higher-dimension operators~\cite{Kusenko:2005du,Kasuya:2014ofa,Cotner:2016dhw,Kawasaki:2005xc}. We parametrize the total model-dependent Q-ball decay width as
	\begin{equation}
	    \Gamma_Q = \dfrac{1}{\tau_Q}~,
	\end{equation}
	where $\tau_Q$ denotes the Q-ball lifetime and contributions of all channels are included. Q-balls have found many applications in particle physics, e.g. they play an important role in Affleck-Dine baryogenesis~\cite{Dine:2003ax}.
	
As discussed above, supersymmetry provides a strong motivation for considering PBH formed from the scalar field fragmentation.  In particular, the electroweak-scale supersymmetry naturally results in the PBH masses for which there are no observational constraints and which can account for all dark matter. While different models of supersymmetry-breaking are possible, for definiteness, let us consider gauge-mediated supersymmetry breaking at the scale $M_{\rm SUSY}\sim 100 \ {\rm TeV}$.  At the time of fragmentation,  when the energy density of the condensate is a fraction  $f\sim r_f^{-1/2}\lesssim 1$ of the total energy density ($\rho_\phi \sim M_{\rm SUSY}^4 \equiv f \times \rho_{\rm tot}$), the mass inside the horizon is 
\begin{equation}
    M_{\rm hor}\sim r_f^{-1/4} \Big(\frac{M_{\rm Planck}^3}{M_{\rm SUSY}^2}\Big)\sim 10^{22} {\rm g} \left ( \frac{100~ {\rm TeV}}{M_{\rm SUSY} }\right)^2~. 
\end{equation}

If the number of Q-balls per the horizon is $N_H\sim 100$ and if a typical PBH has a mass consisting of 10 to 100 Q-ball masses, then the mass of a typical PBH is
\begin{equation}
    M_{\rm PBH} \sim r_f^{-1/4} \times  10^{22} {\rm g} \left ( \frac{100~ {\rm TeV}}{M_{\rm SUSY} }\right)^2~,  
\end{equation}
which is consistent with the open window for dark matter in the form of PBH at masses 
\begin{equation}
    10^{17} {\rm g} \lesssim M_{\rm PBH} \lesssim 10^{22} {\rm g}~. 
\end{equation}
The estimates above are crude, and the precise PBH mass function will be calculated below.  

It is intriguing that the supersymmetry just above the electroweak scale predicts the masses of PBH consistent with the current bounds for dark matter. 

\subsubsection{Background energy density of Q-balls}
	
In order to determine the collapse probability of a Q-ball soliton cluster to a BH, we first need to obtain the average background Q-ball density $\braket{\rho_Q(t_f)}$ at fragmentation time $t_f$, which enters into the expression for overdensity  fluctuations $\delta$ of Eq.~\eqref{eq:denfluc}. 
This can be done by finding the total Q-ball mass-energy   within the horizon $M_H = M_H(t_f)$ and then averaging over the horizon volume $V_H = V_H(t_f)$, where all the relevant quantities are evaluated at the fragmentation time. Using previous notation, this corresponds to
	\begin{equation}
	\braket{\rho_Q(t_f)} =   \dfrac{\braket{M_H(Q,N)}_{V_H}}{V_H}   
 = \frac{1}{V_H} \sum_N \int dQ \, M(Q,N) P(Q|V_H) P(N|V_H)~,
	\end{equation}
which approximates $M_H/V_H$ in the large $V_H$ limit. Here, $M(Q,N) = M_Q N = \Lambda |Q|^{\alpha} N$ is the total horizon mass containing $N$ identical Q-balls of mass $M_Q$. We assume that the total field charge $Q_H$ within the horizon is distributed evenly among all the Q-balls and that this charge is conserved, which we enforce through $P(Q|V_H)$. As before, we take  $P(N|V_H)$ to be Poisson distribution, centered around $\overline{N}_H = \overline{n} V_H$ amount of Q-balls present within horizon. Then,
	\begin{align} \label{eq:rhoq}
	\braket{\rho_Q(t_f)} &= \frac{\Lambda}{V_H} \sum_N \int dQ \, \delta(Q-Q_H/N) \, \Big(\frac{(\overline{n} V_H)^N}{N!} e^{-\overline{n}V_H} \Big) |Q|^\alpha N \notag \\
	&= \frac{\Lambda |Q_H|^\alpha}{V_H} \left[ e^{-\overline{N}_H} \sum_{N=0}^{\infty} \frac{1}{N!} 
	\Big(N^{1-\alpha}\Big) \overline{N}_H^N \right] = \frac{\Lambda |Q_H|^\alpha}{V_H} \braket{N^{1-\alpha}} ~.
	\end{align} 
In the very large soliton number limit, the expression in square brackets of Eq.~\eqref{eq:rhoq} approaches $\braket{N^{1-\alpha}} \approx \overline{N}_H^{1-\alpha}$. The energy density then simply becomes 
\begin{equation} \label{eq:rhoqballsimp}
\braket{\rho_Q(t_f)} \approx \frac{\Lambda |Q_H|^\alpha}{V_H} \overline{N}_H^{1-\alpha} = \Lambda \bar{n} \Big|\dfrac{Q_H}{\overline{N}_H}\Big|^{\alpha}~,
\end{equation} 
which is the form we will use in further computations below.

\subsubsection{Cosmological evolution of energy density}

In order to later make connection with observations at present time, we must comment on the cosmological evolution of the energy density and compute the relevant scale factor $a$ after field fragmentation. As a starting point, we assume an initial period of inflation and reheating, resulting in a radiation-dominated epoch with a uniformly distributed charged scalar field condensate as a sub-dominant component of the energy density. At $t_f$, the scalar field fragments into Q-balls, which scale as matter and come to dominate the energy density at a later time $t_Q$. Primordial black holes are produced during the Q-ball matter-dominated era. As the Q-balls decay at $t_R$, radiation comes to dominate again and the resulting PBH density is frozen in and evolves as a non-relativistic matter to present time. The quantities corresponding to these epochs are denoted with associated subscripts. 

\textit{\underline{Initial radiation-dominated era:}}~~~At the end of inflation, the inflaton coherently oscillates at the bottom of potential, resulting in a matter-dominated era. At $t_{\rm RH} = 1/\Gamma_I$ the inflaton decays and reheats the universe, onsetting a radiation-dominated era with temperature of $T_{\rm RH} = 0.55 g_{\ast}^{-1/4} (\Gamma_I M_{\rm pl})^{1/2}$ and radiation energy density
\begin{equation} \label{eq:reheat}
    \rho_R(t_{\rm RH}) = \dfrac{\pi^2}{30}g_{\ast}(T_{\rm RH}) T_{\rm RH}^4 \approx \dfrac{\pi^2}{327} \Gamma_I^2 M_{\rm pl}^2~\approx \dfrac{\pi^2}{327} 
    \dfrac{M_{\rm pl}^2}{t_{\rm RH}^2} ~,
\end{equation}
where $\Gamma_I \sim 1/t_{\rm RH}$ is the decay width of the inflaton and $g_{\ast}(T)$ denotes the relevant number of temperature-dependent relativistic degrees of freedom. During this period, using Eq.~\eqref{eq:reheat}, the radiation redshifts as
\begin{equation}
    \rho_R (t) = \rho_R(t_{\rm RH}) \Big(\dfrac{a_{\rm RH}}{a}\Big)^{4} = \rho_R(t_{\rm RH}) \Big(\dfrac{t_{\rm RH}}{t}\Big)^{2} \approx \dfrac{\pi^2 M_{\rm pl}^2}{327 t^2} ~~~~~ \text{era:}~~t_{\rm RH} < t < t_Q
\end{equation}

At $t_f$, the scalar field condensate fragments into Q-balls, with the energy density given by Eq.~\eqref{eq:rhoqballsimp}. The time evolution of Q-balls follows that of decaying non-relativistic matter
\begin{align}
    \braket{\rho_Q(t)} =&~ \braket{\rho_Q(t_f)} \Big(\dfrac{a_f}{a}\Big)^3 e^{-(t-t_f)/\tau_Q} \notag \\
    =&~ \dfrac{3 \Lambda Q_f^{\alpha} N_f^{1-\alpha}}{4 \pi t_f^{3/2}t^{3/2}} e^{-(t-t_f)/\tau_Q} ~~~~~ \text{era:}~~t_f < t < t_Q
\end{align}

\textit{\underline{Q-ball matter-dominated era:}}~~~As Q-balls evolve as matter, at $t_Q$ they will come to dominate the energy density, with $\rho_R (t_Q) = \braket{\rho_Q(t_Q)}$. Since Q-balls decay, the temperature associated with radiation will decrease less slowly compared to expansion alone. Following \cite{Scherrer:1985}, the radiation density in the presence of reheating decays is given by
 \begin{gather}
	\rho_R(t) = \left[ \rho_R(t_Q) + \braket{\rho_Q(t_Q)} \int_{x_0}^{x} dx^\prime z(x^\prime) e^{-x^\prime} \right] z^{-4}~,
	\end{gather}
	where $x = \Gamma_Q t$, $x_0 = \Gamma_Q t_Q$  and $z = (x/x_0)^{2/3}$. The Q-balls continue to redshift as well as decay, with the energy density of
	\begin{gather}
	\braket{\rho_Q(t)} = \frac{3 \Lambda Q_H^\alpha \overline{N}_H^{1-\alpha} t_Q^{1/2}}{4\pi t_f^{3/2} t^2} e^{-(t-t_f)/\tau_Q}  ~~~~~ \text{era:}~~t_Q < t < t_R
	\end{gather}
	
	With Q-balls decaying, radiation eventually comes to dominate the universe again, defined by $\rho_R(t_R) = \braket{\rho_Q(t_R)}$. At $t_R$, the radiation density is given by
	\begin{align}
	    \rho_R(t_R) &= \rho_R(t_Q) \left[ 1 + \int_{x_0}^{x} dx^\prime z(x^\prime) e^{-x^\prime} \right] z^{-4} \notag\\
	    &= \frac{M_{\rm pl}^2}{327 t_Q^2} \left[ 1 + \left(\frac{t_Q}{\tau_Q}\right)^{-2/3} \Gamma\left( \frac{5}{3}, \frac{t_Q}{\tau_Q}, \frac{t_R}{\tau_Q} \right) \right] \left(\frac{t_R}{t_Q}\right)^{-8/3}~,
	\end{align}
	where $\Gamma(n,x_0,x) = \int_{x_0}^x dy\, y^{n-1} e^{-y}$ is the generalized incomplete gamma function.
	Equating this to the Q-ball density $\braket{\rho_Q(t_R)}$, one has
	\begin{gather} \label{eq:Constraint}
	1 + \left(\frac{t_R}{\tau_Q}\right)^{-2/3} \Gamma\left(\frac{5}{3}, \frac{t_Q}{\tau_Q}, \frac{t_R}{\tau_Q}\right) = \left(\frac{t_R}{t_Q}\right)^{2/3} e^{(t_Q - t_R)/\tau_Q}~,
	\end{gather}
	which can be restated in terms of $r = t_R/t_Q$ and $r_Q = \tau_Q/t_Q$ as
		\begin{gather}
	\left[1 + \left(\frac{r}{r_Q}\right)^{-2/3} \Gamma\left(\frac{5}{3}, \frac{1}{r_Q}, \frac{r}{r_Q}\right)\right] r^{-2/3} e^{(r-1)/r_Q} = 1~. \label{eq:MoreStableConstraint}
	\end{gather}
	We can now solve numerically\footnote{We find that Eq.~\eqref{eq:MoreStableConstraint} produces more numerically stable results than Eq.~\eqref{eq:Constraint}, while also requiring specification of one less variable.} for $r_Q$ as a function of $r$. Specifying ($r_f = t_Q/t_f$, $t_f$, $r$), allows us to calculate the rest of the parameters $t_Q$, $t_R$, $\tau_Q$ and $\Lambda Q_H^\alpha$ via
	\begin{gather}  
	t_Q = t_f r_f ~~~~~,~~~~~   t_R = t_f r_f r ~~~~~,~~~~~ \tau_Q = t_f r_f r_Q(r) \notag\\ \label{eq:derivfunc}
	\Lambda Q_H^\alpha = \frac{4\pi M_{\rm pl}^2 t_f}{981 \, r_f^{1/2} \overline{N}_H^{1-\alpha}} e^{(1 - 1/r_f)/r_Q(r)}
	\end{gather}
	from which we can calculate all other quantities.
	
\textit{\underline{Standard cosmological era:}}~~~After Q-ball decay, return to the radiation era onsets the standard cosmology. The additional extended early-time matter-dominated era shifts the time at which the standard cosmological events (BBN, matter-radiation equality or recombination) take place and evolution of PBHs and radiation should be determined based on thermal history, where cosmological events occur at a specific temperature. Then, the ratio of two scale factors is given by
	\begin{gather}
	\frac{a_1}{a_2} = \left(\frac{g_{*}(T_2)}{g_{*}(T_1)}\right)^{1/4} \frac{T_2}{T_1}~.
	\end{gather}
	We consider the evolution  from the era at the end of the matter-dominated epoch, with temperature $T_R$ (defined at $\rho_R(t_R)$, following Eq.~\eqref{eq:reheat}),  to the present epoch with temperature $T_0 = 2.7 \text{ K} = 2.3$ meV. To avoid entropy injection during BBN, we require $T_R > T_{\rm BBN} \sim $ MeV. The resulting present-day scale factor relevant for PBH is then given by
	\begin{gather} \label{eq:scalefacqball}
	a = \left(\frac{t_Q}{t_f}\right)^{1/2} \left(\frac{t_R}{t_Q}\right)^{2/3} \left(\frac{g_{*}(T_R)}{g_{*}(T_0)}\right)^{1/4} \frac{T_R}{T_0}~.
	\end{gather}
	In above, $a(t_f) = 1$ has been implicitly assumed.
	
	\subsubsection{PBH mass function}
	Using the formalism of Section~\ref{sec:pbhform}, the redshifted PBH mass function is obtained from evaluation of 
	\begin{align}\label{eq:pbhqballmain}
	\dd{\braket{\rho_\text{PBH}}}{M} &= \dfrac{1}{a^3}   \int \dfrac{dV}{V^2}  \, P(M|V)   B(M,V) M~.
	\end{align}
   The probability that a soliton cluster of mass $M$ will result in a BH is given by~\cite{Kodama:1986ud}\footnote{In our earlier studies~\cite{Cotner:2016cvr,Cotner:2017tir}, for concreteness, we have assumed that BH collapse probability in matter-dominated era follows the analyses of~Ref.~\cite{Polnarev:1986bi,Kokubu:2018fxy}. This does not fully capture the behavior of sub-horizon isocurvature perturbations present in our scenario.~Further, we have previously assumed that overdensities will grow in the matter-dominated era. However, at the linear level of analysis, no overdensity growth is expected. It is a non-trivial task to properly capture analytically the non-linear behavior of the perturbations and we thus omit here this assumption.}
	\begin{gather} \label{eq:probcollapse}
		B(M,V) = K \theta\Big[\delta_0 \Big(\dfrac{M}{M_H r_f}\Big)^{1/3} - \delta_c\Big]~,
	\end{gather}
	where the initial overdensity fluctuation is given by
	\begin{equation}
\delta_0 (M,V) = \dfrac{M}{V\braket{\rho_Q(t_f)}}-1~
	\end{equation}
	and the horizon mass at the beginning of matter/Q-ball domination within horizon volume $V_Q = V_H(t_Q)$ is
	\begin{equation}
M_Q = M_H(t_Q) =  \braket{\rho_Q(t_Q)} V_Q = M_H \bigg(\dfrac{t_Q}{t_f}\bigg)^{3/2} e^{-(t_Q-t_f)/\tau_Q}~.
	\end{equation} 
	The step function $\theta$ selects regions where the overdensity at Q-ball matter-dominated era is larger than the critical overdensity threshold $\delta_c \sim \mathcal{O}(0.1)$ necessary for a black hole collapse. Additional features of a soliton collection such as asphericity, inhomogeneity and angular momentum~\cite{Kokubu:2018fxy,Harada:2017fjm,Harada:2017fjm} will affect the collapse and the associated BH formation probability. However, a detailed study of how these considerations will affect our scenario is beyond the scope of this work. We parametrize these effects in Eq.~\eqref{eq:probcollapse} by a phenomenological prefactor $K$.
The probability of collapse as given by Eq.~\eqref{eq:probcollapse} is further subject to additional constraints, such as those of Section~\ref{sec:NecessaryConditions}, which can be translated into effective limits on the allowed range of values for soliton number $N$ and charge $Q$ that will lead to the formation of black holes.  
	
	The probability density of soliton clusters in Eq.~\eqref{eq:pbhqballmain} can be further decomposed as
	\begin{align} \label{eq:qballmv}
	P(M|V) &= \sum_N \int dQ\, P(M|N,Q)P(Q|N,V)P(N|V) \notag\\
	&= \sum_N \int dQ\, \delta\left(M - \Lambda |Q|^\alpha N \right)  \delta\left(Q - \frac{Q_H V}{N V_H}\right) \left[ \frac{(\bar{n}V)^N}{N!} e^{-\bar{n}V} \right] \notag\\
	&= \sum_N \delta\left(M - \dfrac{\Lambda Q_H^\alpha N^{1-\alpha} V^\alpha}{V_H^\alpha}\right) \left[ \frac{(\bar{n}V)^N}{N!} e^{-\bar{n}V} \right]~,
	\end{align}
	where in the last step we have integrated out the delta-function associated with Q-ball charge as $P(M|N,Q)$ is the only part of the integral that depends on $Q$. The indirect dependence of $B(M,V)$ on $Q$ will be also automatically accounted for. In the above we have assumed that all Q-balls within a cluster are identical and that all clusters have the same overall charge density. 
	
	Combining the above, the resulting PBH mass function contains a delta-function restricting volumes $V$
	\begin{align} \label{eq:pbhqballorig}
	a^3 \dd{\braket{\rho_\text{PBH}}}{M} &=  
	\sum_N \int \dfrac{dV}{V^2}  \, \delta\left(M - \dfrac{\Lambda Q_H^\alpha N^{1-\alpha} V^\alpha}{V_H^\alpha}\right) \left[ \frac{(\bar{n}V)^N}{N!} e^{-\bar{n}V} \right] M\notag\\
	&\quad\times K \theta\Big[\delta_0 \Big(\dfrac{M}{M_H r_f}\Big)^{1/3} - \delta_c\Big]~.
	\end{align} 
	To ensure that the volume set by the delta-function is within the integration range, we require that
	\begin{equation} \label{eq:vconst}
	   V_{\rm min} \, < \frac{V_H}{Q_H} \left(\frac{M}{\Lambda N^{1-\alpha}}\right)^{1/\alpha} < \,
	  V_{\rm max}~, 
	\end{equation}
	with
	\begin{equation}
V_{\rm min} = \dfrac{V_H}{\overline{N}_H} ~~~~~\text{and}~~~~~  V_{\rm max} = V_H \times \text{min}\left[1,   \dfrac{M/M_H}{(1+\delta_c (M_H r_f/M)^{1/3})}\right]~~.
	\end{equation}
	The lower bound $V_{\rm min}$ sets the smallest volume that could collapse and form a PBH, which we take to be on the order of the volume occupied by a single Q-ball. The upper bound $V_{\rm max}$ is the smaller of either a) the horizon volume at fragmentation (since this sets the max length scale of density perturbations by causality), or b) the largest volume of a cluster (given fixed mass) that will collapse to form PBH. The latter originates from trading the step-function for a limit on $V_{\rm max}$, found from using the definition of $\delta_0$ to solve for $V$. This can be intuitively understood as a maximum volume where the total overdensity is larger than the average density within that volume, as increasing it will dilute the density below the threshold.
	
Using the obtained scale factor from Eq.~\eqref{eq:scalefacqball}, the present-day differential fraction of dark matter in PBH is given by 
	 \begin{empheq}[box=\tcbhighmath]{align}  
  \dd{f_\text{DM}}{M} =&~ \frac{1}{a(t_0)^3} \frac{1}{\rho_\text{DM}} \left.\dd{\braket{\rho_{\rm PBH}}}{M}\right|_{t=t_f} \notag\\ \label{eq:qballpbhfinal}  
 =&~\frac{1}{\rho_\text{DM}} \left[ r_f^{1/2} r^{2/3}
  \left(\frac{g_{*}(T_R)}{g_{*}(T_0)}\right)^{1/4} \frac{T_R}{T_0}\right]^{-3}    \\   
	&\times   
	\sum_N \int \dfrac{dV}{V^2}  \, \delta\left(M - \dfrac{\Lambda Q_H^\alpha N^{1-\alpha} V^\alpha}{V_H^\alpha}\right) \left[ \frac{(\bar{n}V)^N}{N!} e^{-\bar{n}V} \right] M\notag\\
	&\times K \theta\Big[\delta_0 \Big(\dfrac{M}{M_H r_f}\Big)^{1/3} - \delta_c\Big] \notag  
	\end{empheq}
	where $r_f =t_Q/t_f$ denotes ratio of time of the beginning of the Q-ball matter-dominated era $t_Q$ and the time at fragmentation $t_f$.
 
\subsubsection{Conditions for collapse to a black hole}

\begin{figure}[ht]
\centering
\includegraphics[width=1.0\linewidth]{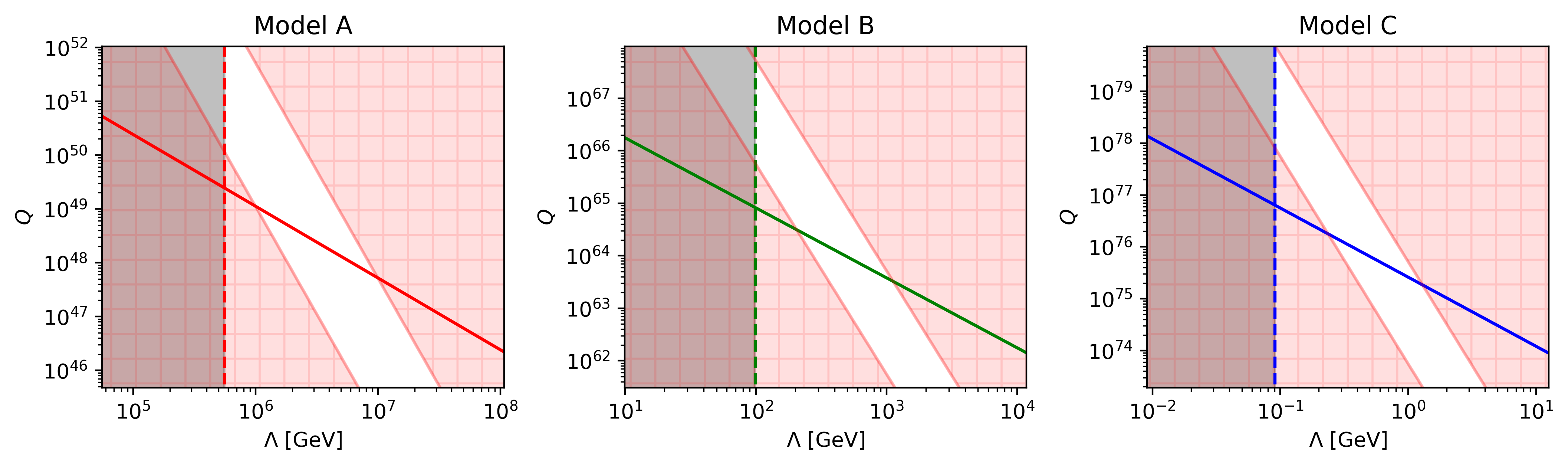}
\caption{Constraints on the model A (left), B (center), C (right) input parameters $ \Lambda, Q$ from the BH collapse conditions 1 and 2 from Section \ref{sec:NecessaryConditions}. Solid colored lines are isocontours of constant $M_\text{Q-ball} = \Lambda Q^\alpha$ and valid values of $\Lambda$ and $Q$ lie on this line. Red hatched regions are excluded due to Eq.~\eqref{eq:qballconst1} and apply equally to all models. Dark regions to the left of the dashed  colored lines are excluded due to restriction of Eq.~\eqref{eq:bhcondqball2}.}
		\label{fig:qballsconstraints}
\end{figure}
	
	\begin{table}
	    \centering
    	\begin{tabular}{|c||c|c|c|c|c|c|}
    	    \hline
    	    Model & $t_f$ [s] & $r_f$ & $r$ & $\overline{N}_H$ & $\alpha$ & $\epsilon$\\ \hline\hline
    	    A & $3\times 10^{-16}$ & 2 & $100$ & $100$ & $3/4$ & 1 \\
    	    B & $7\times 10^{-9}$ & 2 & $40$ & $30$ & $3/4$ & 1 \\
    	    C & $4\times 10^{-3}$ & 1.1 & $30$ & $30$ & $3/4$ & 1 \\ \hline
    	\end{tabular}
    	\caption{Model input parameters for production of PBHs from clustering of Q-balls. As explained in the text, Model A is consistent with electroweak-scale supersymmetry.}
    	\label{tab:qball_model_parameters}
	\end{table}
	
	\begin{table}[]
	    \centering
	    \begin{tabular}{|c||c|c|c|c|c|}
	        \hline
	          &   & $T_R$  & $M_\text{Q-ball}$  & $M_\text{PBH,peak}$  & $M_\text{PBH,peak}$  \\
	         Model & $f_\text{DM}$ &  [MeV] &  [GeV] &  [g] &  [M$_\odot$] \\
	        \hline\hline
	        A & 1.0 &$ 9.0\times 10^{5}$ & $6.1\times 10^{42}$ & $7.8\times 10^{20}$ & $3.9\times 10^{-13}$ \\
	        B & $1.2\times 10^{-2}$ & $3.4\times 10^{2}$ & $4.8\times 10^{50}$ & $1.7\times 10^{28}$ & $8.4\times 10^{-6}$ \\
	        C & $8.5\times 10^{-4}$ & $1.1$ & $3.6\times 10^{56}$ & $6.3\times 10^{33}$ & $3.1$ \\ \hline
	    \end{tabular}
	    \caption{Derived quantities for models A, B, C described in Table \ref{tab:qball_model_parameters}.}
	    \label{tab:qball_derived_quantities}
	\end{table}

In deriving the final expression for PBH mass-spectrum Eq.~\eqref{eq:qballpbhfinal}, the model-specific quantities, energy scale $\Lambda$ of the potential and charge of a single Q-ball $Q$ have been taken as free   parameters and only the product $M_\text{Q-ball} = \Lambda Q^\alpha$ is fixed (see Eq.~\eqref{eq:derivfunc}). The BH collapse restrictions of Section~\ref{sec:NecessaryConditions} will constrain the allowed $\Lambda$ and $Q$ parameter space. 
	
As an illustration, we show how these restrictions affect the parameter space of several concerete models using computations of Ref.~\cite{Cotner:2016cvr,Cotner:2017tir} and input parameters given in Table~\ref{tab:qball_model_parameters}, with their associated derived quantities shown in Table~\ref{tab:qball_derived_quantities}.
In particular, by using the definitions of $M$, $M_H$ and Eq.~\eqref{eq:rhoqballsimp},  Eq.~\eqref{eq:bhcondqball1} can be recast as a constraint on $N$ 
	\begin{equation} \label{eq:qballconst1}
		1 < \dfrac{1}{2} \left[ \frac{M_H N}{M Q_H^\alpha \overline{N}_H^{1-\alpha}} \right]^\frac{\alpha-\beta}{\alpha} \left(\frac{M_{\rm pl}}{\Lambda}\right)^2 < N^{2/3}~.
	\end{equation}
The above is further simplified once we consider the special case of Q-balls from gauge-mediated SUSY, with $\alpha=3/4$ and $\beta=1/4$, which sets $(\alpha-\beta)/\alpha = 1$. The other BH formation condition of Eq.~\eqref{eq:bhcondqball2} can be viewed as simply a restriction on $\Lambda$. The result of imposing these constraints is shown in Figure~\ref{fig:qballsconstraints}, where each line corresponds to model A, B, C depicted on the constraints plot on the left.
	
\section{Discussion and conclusion}
\label{sec:sum}

In this work we have developed a general analytic description of PBH formation from scalar field fragmentation in the early universe. 

Formation of PBHs from scalar field dynamics appears to be very common.  In addition to the Higgs field, whose existence has been decisively confirmed, theories beyond the Standard Model generally predict appearance of multiple scalar fields.~Any one of these fields can develop a large expectation value and undergo fragmentation under some very mild constraints on the shape of the potential. More specifically, the potential should be shallow enough to allow for a large VEV at the end of inflation, in accordance with Eq.~\eqref{eq:initial_VEV}, and the second derivative of the potential should be small or negative, in accordance with Eq.~\eqref{eq:instcond}.  A good example is supersymmetry. All supersymmetric generalizations of the Standard Model contain multiple ``flat directions" along which the potential is zero in the limit of exact supersymmetry and which are lifted by supersymmetry breaking terms~\cite{Gherghetta:1995dv}.  The soft supersymmetry breaking terms usually result in a flat direction that grows slowly at large VEV, so that it develops a large VEV at the end of inflation.  In the case of gauge-mediated supersymmetry breaking, the condition of Eq.~\eqref{eq:instcond} is also satisfied at any scale above the supersymmetry breaking scale.  Therefore, such flat directions are well suited for producing PBHs.

Given the preponderance of scalar fields with the requisite properties in theories beyond the Standard Model, PBH formation appears to be a fairly generic phenomenon. This makes PBHs an appealing candidate for dark matter in theories with supersymmetry or other theories involving scalar fields. 

\acknowledgments
\addcontentsline{toc}{section}{Acknowledgments}
 
The work of E.C., A.K. and V.T. was supported by the U.S. Department of Energy (DOE) Grant No. DE-SC0009937.~The work of M.S. was supported by the Ministry of Education, Culture, Sports, Science and Technology of Japan (MEXT) and by Japan Society for the Promotion of Science (JSPS) KAKENHI Grants No.~15H05888 and No.~15K21733. A.K. and M.S. were also supported by the World Premier International Research Center Initiative (WPI), MEXT, Japan. Part of this work was performed at the Aspen Center for Physics, which is supported by National Science Foundation grant PHY-1607611.

\clearpage
\bibliography{pbhfragref}
\addcontentsline{toc}{section}{Bibliography}
\bibliographystyle{JHEP}
\end{document}